\documentclass[10pt, twocolumn, letter]{paper}
\pdfoutput=1

\usepackage[pdftex]{graphicx}
\graphicspath{{../images/}}
\DeclareGraphicsExtensions{.pdf,.jpeg,.png,.jpg}
\usepackage{array}

\usepackage{amssymb}
\usepackage{amsmath}

\usepackage{url}

\usepackage[tight,footnotesize]{subfigure}

\newcommand{\mmmks}{3MK Simulations}

\begin{document}
\title{High-Performance Distributed Multi-Model / Multi-Kernel Simulations: A Case-Study in Jungle Computing}

\author{
Niels Drost\hspace{0.5cm}Jason Maassen\hspace{0.5cm}Maarten A.J. van Meersbergen\hspace{0.5cm}Henri E. Bal\\
\vspace{0.3cm}\small Dept. of Computer Science, VU University, Amsterdam, The Netherlands, Email: niels, jason, maarten, bal@cs.vu.nl\\
F. Inti Pelupessy\hspace{0.5cm}Simon Portegies Zwart\\
\vspace{0.3cm}\small Leiden Observatory, Leiden University, Leiden, The Netherlands. Email: pelupes, spz@strw.leidenuniv.nl\\
Michael Kliphuis\hspace{0.5cm}Henk A. Dijkstra\\
\small Institute for Marine and Atmospheric research Utrecht,
Department of Physics and Astronomy,\\
\vspace{0.3cm}\small Utrecht University, Utrecht, The Netherlands. Email: m.kliphuis, h.a.dijkstra@uu.nl\\
Frank J. Seinstra\\
\small Dept. of Computer Science, VU University, Amsterdam, The Netherlands, Email: fjseins@cs.vu.nl\\
} 


\maketitle

\begin{abstract}
High-performance scientific applications require more and more compute power.
The concurrent use of multiple distributed compute resources is vital for making scientific progress.
The resulting distributed system, a so-called \emph{Jungle Computing System}, is both highly 
heterogeneous and hierarchical, potentially consisting of grids, clouds, stand-alone machines, clusters, desktop grids, mobile devices, and supercomputers, possibly with accelerators such as GPUs.

One striking example of applications that can benefit greatly of Jungle Computing Systems are Multi-Model / Multi-Kernel simulations.
In these simulations, multiple models, possibly implemented using different techniques and programming models, are coupled into a single simulation of a physical system.
Examples include the domain of computational astrophysics and climate modeling.

In this paper we investigate the use of Jungle Computing Systems for such Multi-Model / Multi-Kernel simulations.
We make use of the software developed in the \emph{Ibis} project, which addresses many of the problems faced when running applications on Jungle Computing Systems.
We create a prototype Jungle-aware version of \emph{AMUSE}, an astrophysical simulation framework.
We show preliminary experiments with the resulting system, using clusters, grids, stand-alone machines, and GPUs.
\end{abstract}


\section{Introduction}

\begin{figure*}
\centering
\includegraphics[width=\textwidth]{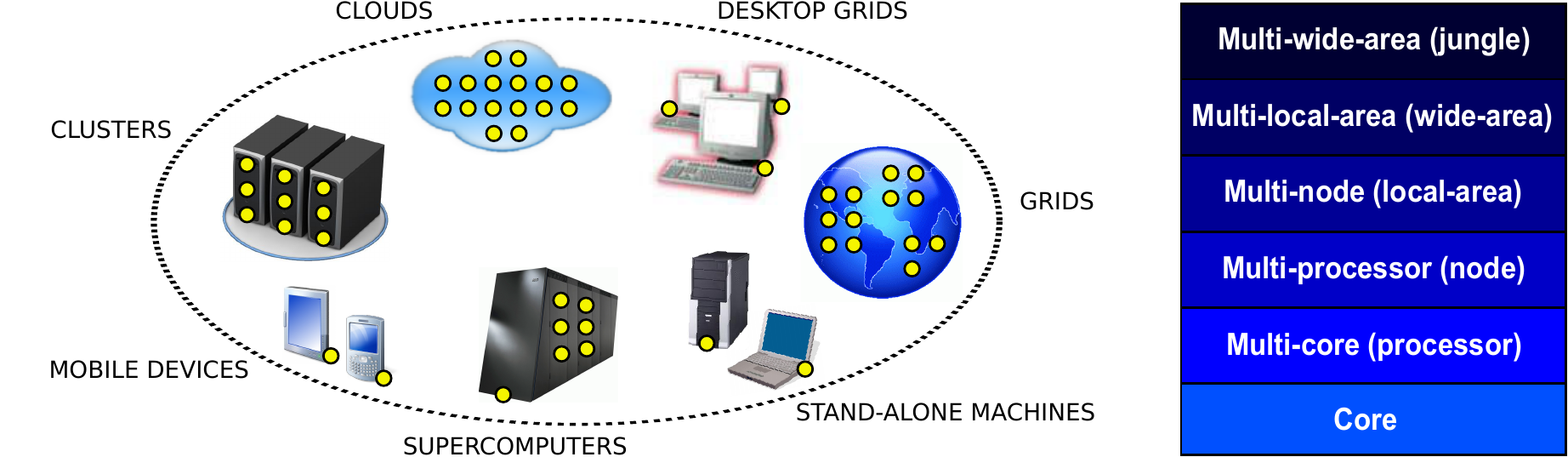}
\caption{Left: A worst case 'Jungle Computing System', a distributed computing platform, consisting of clusters, clouds, grids, desktop grids, supercomputers, as well as stand-alone machines and possibly even mobile devices. All resources may include accelerators such as GPUs and FPGAs. Users are forced to use such complex systems to perform single computations concurrently. Typically, an end-user uses some subset of all resources at any time.
Right: A Jungle Computing System is highly hierarchical, with possibilities for parallelism at all levels. This image taken from~\cite{jungle}}
\label{figure_jungle}
\end{figure*}

The high-performance distributed computing landscape is increasingly complex, consisting of clusters, grids, clouds, supercomputers, and other platforms.
Moreover, accelerators such as GPUs are quickly becoming the norm, adding to the complexity even further.
Because of the sheer size of scientific problems, many scientists are forced to make use of multiple different resources, 
and combine these in a single run of an application.
Combining resources may be necessary if no single resource is available that is large enough to perform the required computation, 
or because different parts of the computation have different computational requirements.
We call such concurrent usage of heterogeneous, hierarchical, and distributed resources \emph{Jungle Computing}~\cite{jungle} (See Figure~\ref{figure_jungle}).

One type of applications where the usage of Jungle Computing is a necessity are Multi-Model / Multi-Kernel simulations, or \mmmks.
In these simulations, multiple models are combined into a single, large simulation of a physical system.
Next to Multi-Model, these simulations are often also Multi-Kernel: for each model, multiple implementations (kernels) may be available.
For example, multiple implementations of a model may exist that generate the same result, but are suitable for different resources (e.g. GPUs vs CPUs).
Also, a kernel may perform a computation faster on the same resource, at the cost of reduced output quality.

One example of such simulations is climate modeling, where models of land, ocean, atmosphere, and ice are combined to simulate the earth's climate 
as a whole~\cite{cesm}. For these simulations, vast amounts of compute power are required, often more than any single supercomputer can supply.
\mmmks{} are also used in computational astrophysics.
In this field, multiple models are used for stellar evolution, hydrodynamics, gravitational dynamics, etc.
In both these domains, running \mmmks{} on a Jungle is vital to make significant steps forward scientifically.

Compute Jungles are both highly heterogeneous and hierarchical, making it hard for scientists to make efficient and effective use of these systems.
For instance, different resources use vastly different middleware to provide access, making it difficult to run an application on a combination of resources.
Also, connectivity between resources is problematic because of firewalls and NATs, among others.
In the Ibis project~\cite{jungle,ibis-ieee}, much work has been done to solve the basic problems present in Jungles.
The resulting software framework\footnote{All Ibis software is open-source and can be downloaded freely from \url{http://www.cs.vu.nl/ibis}} offers 
the basic functionality required to run any application, and thus \mmmks{}, in a Jungle, allowing application developer to focus 
on \emph{problem solving}, rather than \emph{system fighting}.

In this paper, we perform a case study in running \mmmks{} in a Jungle Computing System.
We focus on computational astrophysics, basing our experiments on the AMUSE~\cite{muse} framework\footnote{AMUSE is open-source and can be downloaded freely from \url{http://www.amusecode.org}}.
AMUSE is a software framework for large-scale simulations of dense stellar systems that couples 
existing simulation models, combining them into a single, parallel simulation.
Using the Ibis software framework, we build a prototype distributed version of AMUSE, capable of extending a simulation across a Jungle.

The contributions of this paper are as follows:

\begin{itemize}

\item We describe the general structure of \mmmks{}, and the issues that arise when running these in a Jungle.

\item We show a prototype distributed version of AMUSE that is capable of running in a Jungle.

\item We discuss a number of initial experiments of the resulting system. Some of these spanning compute resources on multiple continents.

\end{itemize}

This paper is structured as follows. In Section~\ref{jungles} we define the Jungle Computing paradigm.
In Section~\ref{ibis} we give a brief overview of the Ibis software framework, which has been explicitly designed for Jungle Computing applications.
In Section~\ref{mmmks} we discuss \mmmks{}, especially focusing on their structure from a computer science perspective. We also describe AMUSE, 
a computational astrophysics framework.
Section~\ref{distributed_amuse} shows how we have added distributed functionality to AMUSE, enabling it for use in the Jungle.
In Section~\ref{evaluation} we evaluate the resulting system with a number of experiments, using a simulation of early star clusters.
Finally, we present some conclusions and future work in Section~\ref{conclusions}.

\section{Jungle Computing}
\label{jungles}

Jungle Computing is a distributed computing paradigm born in practice, rather than from theory.
Instead of being invented, it simply emerged out of the plethora of distributed resources available.
A Jungle Computing System (See Figure~\ref{figure_jungle}) consists of all compute resources available to end-users, including clusters, clouds, grids, desktop grids, supercomputers, as well as stand-alone machines and possibly even mobile devices.

In general, users make use of Jungle Computing Systems out of necessity, rather than choice.
There are several reasons for using Jungle Computing Systems.
Firstly, an application may require more compute power than available in any one system a user has access to.
Secondly, different parts of an application may have different computational requirements, with no single system that meets all requirements.
Finally, long queues and limited availability of resources may lead users to opportunistically choose whatever machine is available 'today', 
or spread their (independent) jobs over multiple resources.

From a high-level view, all resources in a Jungle Computing System are in some way equal, all consisting of some amount of processing power, 
memory, and possibly storage. End-users perceive these resources as just that: a compute resource to run their application on.
Whether this resource is located in a remote cloud or located down the hall in a cluster, is of no interest to an end-user, as long as his or 
her application runs effectively.

Despite this similarity of resources, a Jungle Computing System is highly heterogeneous.
Resources differ in basic properties such as processor architecture, amount of memory, and performance.
As there is no central administration of these unrelated systems, installed software such as compilers and libraries will also differ. 
Moreover, the diverse computing paradigms (cloud, grid, etc) differ in usage model.
For example, where a stand-alone machine is usually permanently available, a grid resource will have to be reserved, while a cloud requires a 
credit card to gain access.
Also, the \emph{middleware} used to access a resource differs greatly, using completely different interfaces.

The heterogeneity of Jungle Computing Systems makes it hard to run applications on multiple resources.
For each used resource, the application may have to be re-compiled, or even partially re-written, to handle the changes in software and hardware available.
Moreover, for each resource, a different middleware interface may be available, requiring different middleware client software.
Once an application has been successfully started in a Jungle (not an easy feat), another aspect that hinders usage of Jungle Computing Systems 
is the lack of connectivity between resources.
Resources, especially clusters and supercomputers, are usually not designed with communication to the outside world in mind, resulting in 
non-routed networks, firewalls, NATs, and other restrictions on communication.

Besides distributed and heterogeneous, Jungle Computing Systems are also hierarchical, as shown in Figure~\ref{figure_jungle}.
There are many possibilities for concurrency and parallelism that must be exploited by applications to make efficient use of all resources.

\section{The Ibis Software Framework}
\label{ibis}

As the previous section shows, a large number of problems must be addressed to make successful use of Jungle Computing Systems.
In the Ibis~\cite{jungle,ibis-ieee} software framework, we have implemented solutions that together meet the basic requirements for running applications 
in such systems.
The Ibis project has been running for ten years, giving us considerable experience with running applications in these 'wild' environments.

\begin{figure}
\centering
\includegraphics[angle=-90,width=\columnwidth]{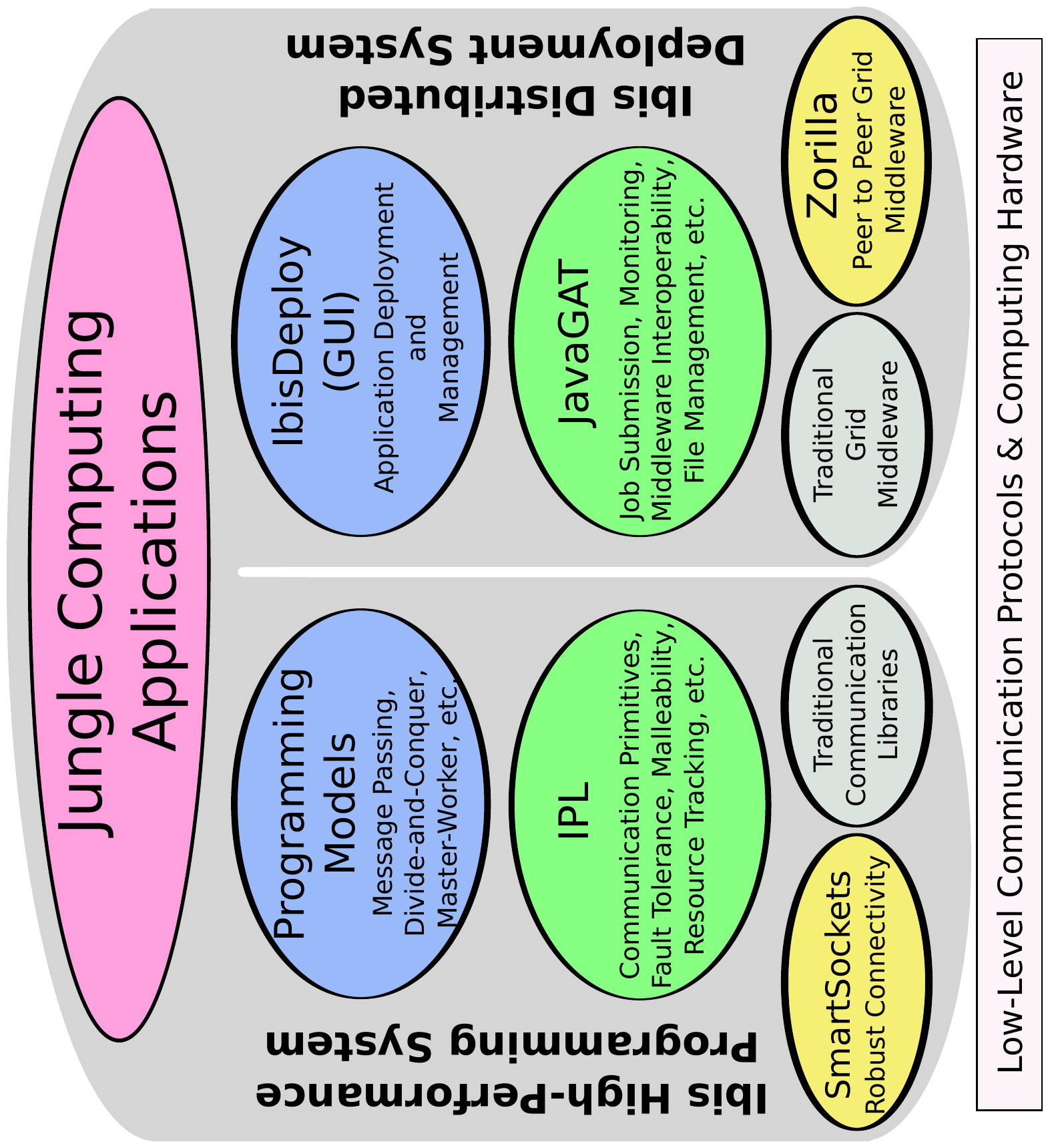}
\caption{Overview of the Ibis Software Framework, consisting of functionality for programming application on the left, and functionality for deploying applications in the Jungle on the right.}
\label{figure_ibis_design}
\end{figure}

Ibis is designed in a modular way, and it is possible to use each part separately.
Users can choose to use as few or as many of the parts as they desire.
Ibis can easily interface with existing software, written in any programming language.
In this section, we will give a brief description of each of the parts.
For more information see~\cite{jungle} and~\cite{ibis-ieee}.

Figure~\ref{figure_ibis_design} gives an overview of the Ibis software framework that consists of two distinct subsystems.
The \emph{Ibis Distributed Deployment System} shown on the right allows users to start applications in the Jungle with ease.
The central component in this subsystem is JavaGAT~\cite{javagat}.
JavaGAT is a generic and simple interface to middleware. Instead of writing software for one specific middleware, thereby limiting its use, applications can use the generic JavaGAT interface.

Using familiar concepts such as \emph{Files} and \emph{Jobs}, a programmer is able to start applications in a Jungle.
JavaGAT provides this functionality using \emph{Adapters}, that interact with a middleware to implement the required task, be it copying a file, 
starting a job, monitoring a system, or otherwise.
JavaGAT will automatically select the appropriate adapter for each resource, and adapters exist for most common middleware including Globus, 
Unicore, SSH, Glite, SGE, PBS.
Adapters are easily added if needed. JavaGAT users regularly contribute new adapters to the main JavaGAT distribution.

It is important to note that although JavaGAT, as all software in the Ibis framework, is written in Java, it supports running any application, written in C, C++, Fortran, MPI, or otherwise.

Besides interfacing with existing middleware, JavaGAT is also able to use Zorilla~\cite{zorilla}, a prototype middleware based on Peer-to-Peer techniques.
Zorilla is ideal in cases where no middleware is available, and can turn any collection of machines into a cluster-like system in minutes.

In addition to JavaGAT, Ibis also provides \emph{IbisDeploy}: a library for deploying application in the 
Jungle, targeted specifically at end-users.
IbisDeploy can be configured using a small number of simple configuration files, or with an optional 
GUI (see Figures~\ref{figure_deploy_screenshot_jobs} and~\ref{figure_deploy_screenshot_network_viz}, and Section~\ref{evaluation}).

The left of Figure~\ref{figure_ibis_design} shows the \emph{Ibis High-Performance Programming System}.
In contrast to the previously discussed subsystem which allows users to deploy any application in the Jungle, functionality provided by this 
part of the Ibis framework allows programmers to write applications specifically designed to run in a Jungle Computing System.
We will discuss two of the three parts of this subsystem here: IPL, and SmartSockets.
The last, the \emph{Programming Models} present in Ibis are not used by the software described in this paper, and are discussed 
elsewhere~\cite{jungle, ibis-ieee}.

The central component in this part of Ibis is the \emph{Ibis Portability Layer}, or IPL~\cite{ipl}.
IPL is a communication library specifically designed for use in a Jungle.
IPL is based on the concept of uni-directional connection-oriented message-based communication.
It provides support for fault-tolerance and malleability, two things required to run successfully in a Jungle environment.
For instance, an application using IPL will get notified if a machine crashes, allowing the application to react to and recover from this fault.
Like JavaGAT, IPL is mainly an API, and functionality is provided by multiple implementations on top of an existing communication library, with the best automatically selected at runtime.
Some implementations are implemented in Java entirely, resulting in excellent portability, while other use C-code to provide extra performance when special-purpose networks are available.
A number of implementations exist, including interfaces to TCP, Infiniband, Bluetooth, and Myrinet.

To combat the connectivity problems present in Jungle Computing Systems, Ibis provides functionality for dealing with firewalls, NATs, non-routed networks, and such.
SmartSockets~\cite{smartsockets} provides a socket-like interface, while automatically dealing with any communication problems.
For this, SmartSockets uses an \emph{overlay} network, consisting of a number of \emph{hubs}. These hubs typically run on machines with more connectivity, such as the front-end machine of a cluster.

The overlay network provides a way to coordinate communication, and serves as a backup communication medium if required.
For instance, firewalls in general only block traffic in one direction, refusing incoming connections, while still allowing outgoing traffic.
As a result, a connection setup to a machine behind a firewall is likely to fail. However, when using SmartSockets, the overlay network 
can be used to send a 'reverse connection request' to the target machine. This machine can then create an outgoing connection, thereby 
circumventing the firewall.

Using IPL and SmartSockets together allows applications to easily communicate with all resources used by an application running on a Jungle Computing System.
To make the usage of SmartSockets as easy as possible, IbisDeploy automatically starts the hubs required by SmartSockets on each resource used.

As explained above, Ibis can integrate existing software written in C, Fortran, or any other language.
In this case, an application uses IPL to communicate across a Jungle, and interfaces to existing software performing the actual computation.
This way, Ibis is used as \emph{glue} between the parts of a computation.

\section{Multi-Model / Multi-Kernel Simulations}
\label{mmmks}

\mmmks are a good example of applications that greatly benefit from the usage of Jungle Computing Systems.
In these simulations, multiple models are combined into a single simulation, for example combining separate atmosphere, ocean, land and ice models
to create a climate modeling simulation.
Moreover, multiple versions of each model (multiple kernels), may be available, for several reasons.
Many models recently have been ported to make use of GPUs now present in many systems.
Which kernel is used (the CPU or the GPU version) has no influence in the result of the simulation, but may have a dramatic effect on performance.
Also, some models can be implemented using different techniques, resulting in different trade-offs in performance versus precision.
Another reason for multiple kernels of a single model is that some implementations of models have a limited scope, and function only within 
certain parameters of the model. Outside of this scope, a different kernel may have to be used.

\mmmks{} combine several models with different performance characteristics, and may perform best on different hardware.
If only a single resource is used, as is currently the case with such simulations, performance may be sub-optimal.
Moreover, the total performance will be limited by the size of the biggest machine available.
As an alternative, using a Jungle Computing System allows each model to be deployed on the best resource.
For instance, any model with a GPU enabled kernel available could be run on a GPU cluster, while other parts of the simulation are run on a supercomputer.
This will increase the performance and efficiency of the simulation, while also increasing the biggest problem size that can be handled.
In scientific simulations the quality of the results generally increases with increasing resolution.
As this in turn leads to a bigger problem size in 
the application, using Jungle Computing Systems directly increases the quality of research that can be performed.

\subsection{Computational Astrophysics: AMUSE}

\begin{figure}
\centering
\includegraphics[angle=-90,width=\columnwidth]{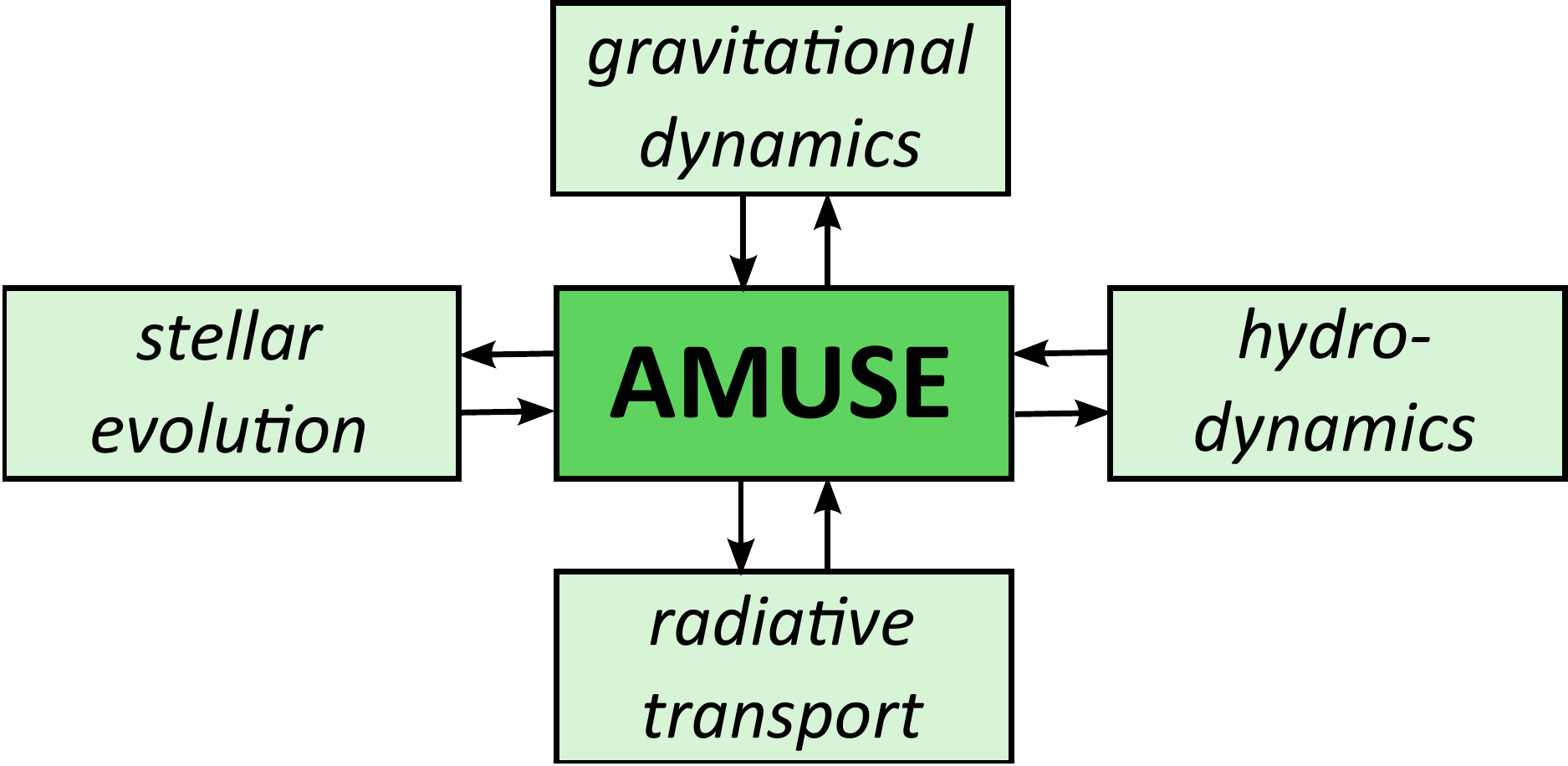}
\caption{High level design of the AMUSE astrophysical simulation framework. AMUSE consist of a central component written in Python, interacting with models written in C, C++, Fortran, CUDA, etc. Communication with these models is done using MPI, sockets, or our new Ibis channel.}
\label{figure_amuse_design}
\end{figure}

We will now discuss two systems using \mmmks{}.
The first is the \emph{Astronomical Multipurpose Software Environment}, or AMUSE~\cite{muse}.
AMUSE is a computational astrophysics framework developed at Leiden Observatory.
As shown in Figure~\ref{figure_amuse_design}, AMUSE combines different models (stellar evolution, hydro-dynamics, gravitational dynamics, 
and radiative transport) into a single astrophysical simulation.
These models may be implemented using any language, and kernels for models exist written using C, C++, and Fortran, combined with CUDA, OpenMP, OpenCL, and MPI.

In AMUSE, models are integrated into a single simulation in a centralized \emph{coupler}.
Like the rest of AMUSE, this coupler is written in Python.
Users write simulations using a high-level API offered by AMUSE.
This API is based as much as possible on the physical interactions of the different types of models, rather than their underlying numerical representation of the physics.
AMUSE implements all functionality required to perform astrophysical simulations, for example by supporting automatic unit conversion.
With the large number of units used in astronomy, checked conversion of all these units is a requirement for combining different models.
Other functionality includes reading and writing data sets in standard formats, and generating initial conditions.
Using Python scripts, an endless number of simulations can be created by combining one or more models offered.
Examples include simulations of supernova explosions, collisions between galaxies, and the orbit of planets in a solar system.
In Section~\ref{evaluation} we will discuss an example simulation: the evolution of embedded star clusters~\cite{stellar_evolution}.

In an AMUSE script, whenever a simulation creates a model, a so-called \emph{worker} is created automatically by the AMUSE runtime.
This worker consists of a model executable running on some resource, offering functionality to the simulation.
AMUSE communicates with workers using a \emph{channel}, in an RPC-like method. Both synchronous and asynchronous calls are supported.
The default channel uses MPI for sending messages to and from a worker, however, a channel based on sockets is also available.
For this paper, we added an \emph{Ibis} channel, which we will discuss in Section~\ref{distributed_amuse}.
All communication required between different models is done through the AMUSE coupler.
This allows proper checking of data transferred, for instance unit conversion, and checking for illegal values and error states.
However, it also introduces a potential bottleneck when large-scale simulations are done.
We regard creating a distributed version of the coupler, or adding direct communication between models as future work.

\subsection{Climate Modeling: CESM}

\begin{figure}
\centering
\includegraphics[angle=-90,width=\columnwidth]{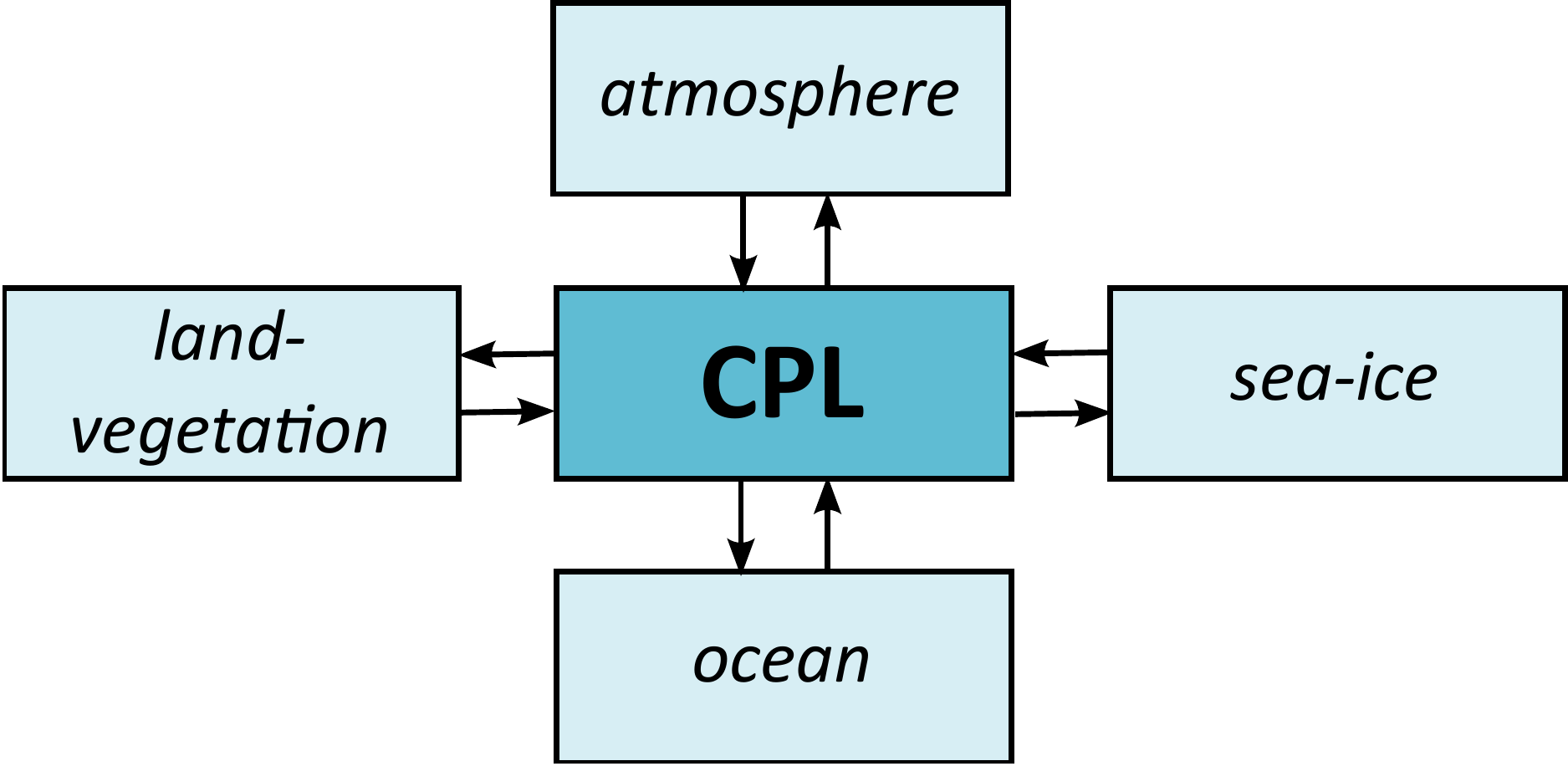}
\caption{High level design of the Community Earth System Model (CESM). In CESM, all models are written in Fortran, MPI, and OpenMP, and are coupled using a parallel coupler also written in Fortran using MPI.}
\label{figure_cpl_design}
\end{figure}

A second example of \mmmks{} is the \emph{Community Earth System Model}~\cite{cesm} or CESM.
CESM couples models for atmosphere, oceans, land and sea-ice into a single simulation of the earth's climate, as is shown 
in Figure~\ref{figure_cpl_design}.

The models are implemented in Fortran, MPI and OpenMP. Various versions of each model exists,
for example, focusing on different eras in Earth's history or using more advanced modeling 
techniques (CAM4 vs. CAM5 atmosphere models). In addition, both \emph{active} and \emph{data} implementations exist of each model. 
The former computes all results, while the latter simply replays precomputed data. 

The design of CESM shown in Figure~\ref{figure_cpl_design} is very similar to the previous AMUSE example.
Unlike AMUSE, however, the central coupler of CESM is designed to run in parallel on (part of) the resources used in a simulation.
Like the models, the coupler is implemented in Fortran and MPI.

CESM is primarily designed to run on one large machine, such as a cluster or supercomputer. 
The application is started as a single MPI job, after which the models are distributed over the available compute nodes according to 
a user defined configuration. The compute nodes can either be \emph{partitioned}, each running (part of) one model, 
\emph{shared}, each running (part of) multiple models, or use a combination of both. 
In this configuration, the coupler is also assigned part of the resources, just like the models. 
Because the computational requirements of each model (and the coupler) vary depending on the experiment, it may take a user quite a bit of experimenting to find an efficient configuration for distributing the models over the available compute nodes. 

To further increase our understanding of \mmmks{}, we are currently in the process of creating a Jungle-aware version of CESM.
We plan to make it possible to run simulations distributed over multiple resources, and provide tools to automatically find an optimal configuration for distributing the models over these resources.
In addition, we are investigating porting parts of the models to GPUs. 

The designs of AMUSE and CESM show a remarkable similarity for such unrelated fields.
Differences between the two systems stem from different implementation approaches, rather than from fundamental differences.
We therefore argue that \mmmks{} are truly a class, rather than an exception.
Insights gained and techniques developed for such systems should be usable in a large number of applications.

\subsection{Requirements of Multi-Model / Multi-Kernel Simulations in Jungle Computing Systems}
\label{requirements}

To successfully run High-Performance Distributed \mmmks{} in a Jungle Computing System, a number of requirements need to be fulfilled.
We list these requirements here.
However, we do not expect this list to be complete, or fulfill all these requirements in this work, 
as we are still very much investigating running these simulations on Jungle Computing Systems.

First, it must be as easy as possible to deploy a simulation.
If it is too complex to start or configure a simulation, this will render it unusable for scientists trying to focus on their research.
This also means that simulations should be flexible: changing a model to a different implementation, or changing a model parameter, should be easy to do.
Input and output files should also automatically be copied to where they are needed.

Second, the application should be able to communicate between all resources.
Communication, and the complete system, should be as fast and efficient as possible.
Having an application run slowly defeats the purpose of Jungle Computing.
The simulation should also scale to large systems, including supercomputers.

Third, it should be possible to do both performance and correctness monitoring of the system.
The bigger the system, the harder it is to oversee.
If a simulation is running inefficiently, or, worse, crashing, a user should be able to find 
the cause without looking at hundreds of log files.

Fourth, it is of vital importance that the software is stable.
A simulation may run for months, and crashes need to be kept to a minimum.
Since a Jungle Computing System is unstable by nature, this requires the software to handle and overcome faults as much as possible.

Fifth and last is a requirement that is high on the wish list of users: the automatic discovery of suitable resources.
Given the list of resources a user has access to, ideally, software should find suitable resources itself, without any intervention from the user.
Likewise, the system should also find replacement resources if some fail, or, more subtle, fail to perform.

\section{Distributed AMUSE}
\label{distributed_amuse}

In this Section we show our \emph{prototype} distributed version of AMUSE.
We are still actively working on this software, and expect to include our Ibis code in the next distribution of AMUSE.

\begin{figure}
\centering
\includegraphics[width=\columnwidth]{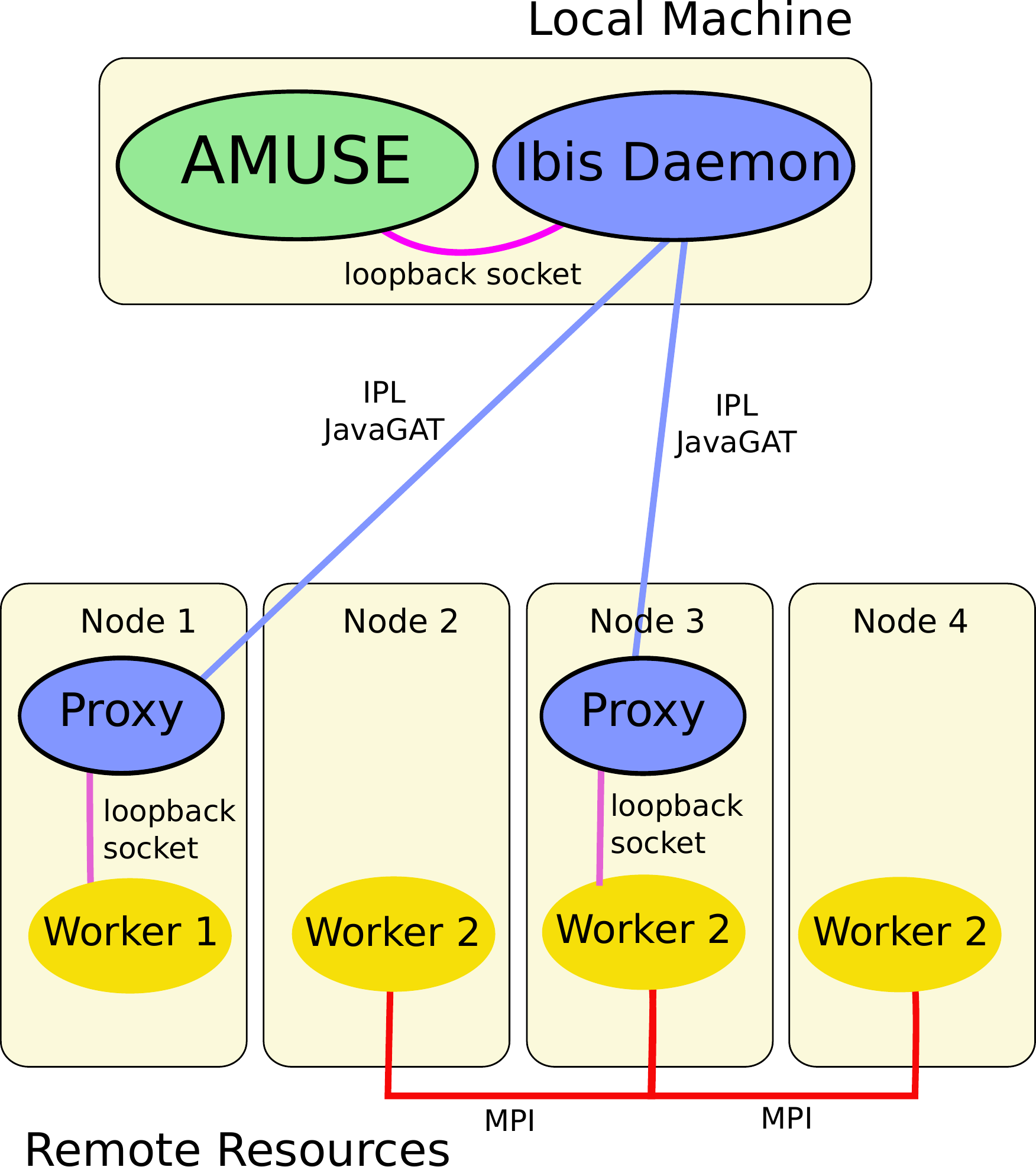}
\caption{Overview of the design of distributed AMUSE, in this case using 4 nodes of a cluster for running 2 different models. The AMUSE coupler connects with a local \emph{Ibis daemon} to start and communicate with remote workers. Workers are started by the daemon with JavaGAT, while wide-area communication is done using IPL. In this case Worker 1 is a sequential model, while Worker 2 uses MPI.}
\label{figure_distributed_amuse}
\end{figure}

To create a version of AMUSE capable of running in a Jungle Computing System we added an \emph{Ibis Channel} to the worker startup and communication code.
Figure~\ref{figure_distributed_amuse} shows the resulting design.  
The AMUSE coupler connects with a local \emph{Ibis daemon} to start and communicate with remote workers.
The user must start this daemon on his or her machine before running any simulation, but it can be re-used for all simulations run.
We use this separate process as the Ibis software is written in Java, while AMUSE is written in Python.
The connection is created using a local loopback socket.
Benchmarks show that this connection is over 8Gbit/second even on a modest laptop, has a extremely small latency, and we expect very little performance issues rising from this extra step in communication.

Once the daemon is running, and a simulation requests a worker to be started, the daemon uses the IbisDeploy library to start the worker on a remote machine.
Our system assumes that AMUSE is already installed on the target resource.
Since AMUSE contains large portions of C, C++, and Fortran, and requires a large number of libraries, installing it automatically would fail on most, if not all, machines.
Fortunately, AMUSE will have to be installed only once per resource, and multiple users can share one installation. Also, a pre-build image could be used on resources such as clouds that employ virtualization.
A JavaGAT job is eventually submitted by IbisDeploy to start the worker process on the remote resource.
As some models are parallel applications, the daemon may start a parallel job, with multiple processes and nodes for a single worker.
IbisDeploy also automatically starts a SmartSockets hub to support communication in case of firewalls and such (not shown in figure).

Once the worker is started the daemon uses IPL to communicate over the wide area connection to a \emph{proxy} process running alongside the worker.
The proxy communicates using a loopback connection with the worker process.
This adds another extra step in communication.
However, a large number of models use MPI internally for parallelism, and mixing Java and MPI in a single process is not advisable.
This would result in instabilities due to the 'tricks' both MPI and Java use, such as overriding the malloc function used for memory allocation.

With these modifications, AMUSE is capable of automatically starting remote workers on any resource the user has access to, without a lot of effort required from the user.
To use the distributed version of AMUSE a user must:

\begin{enumerate}
\item Ensure AMUSE is installed on all resources used.
\item Specify some basic information such as hostname and type of middleware for each resource used in a configuration file.
\item Start the Ibis-Daemon on the local machine.
\item Add a property to each worker created in the simulation script to specify the channel used (ibis), as well as the name of the resource, and the number of nodes required for this worker.
\end{enumerate}

Our prototype system fulfills most of the requirements we specified in Section~\ref{requirements}.
It is easy to use because of IbisDeploy, and communication between all resources is possible thanks to SmartSockets.
Monitoring is possible to a limited degree because of the optional GUI (we will discuss the GUI in Section~\ref{evaluation}).

Stability is still a problem, however.
As long as no faults occur on the resources our system works fine, and our communication library can handle transient network failures without problems.
However, our prototype is not able to handle a machine disappearing very well.
If a reservation ends for a resources, and the worker is killed by the scheduler, we cannot recover from this fault, and the entire simulation crashes.
In theory it should be possible to transparently find a replacement machine, and we see this as future work.

Automatic discovery of resources is another requirement that we do not fulfill.
Although specifying which machine needs to be used is easy, it should not be needed at all.
We have experience with resource discovery in Peer-to-Peer systems~\cite{zorilla}, and plan to use some of this experience in the next version of the Ibis software system.

We conclude that despite its limitations, our prototype distributed AMUSE system is very capable of running \mmmks{} applications in a Jungle.
We performed some initial experiments to verify this, and will show these in the next section.

\section{Evaluation}
\label{evaluation}

\begin{figure}
\subfigure{
\includegraphics[width=\columnwidth]{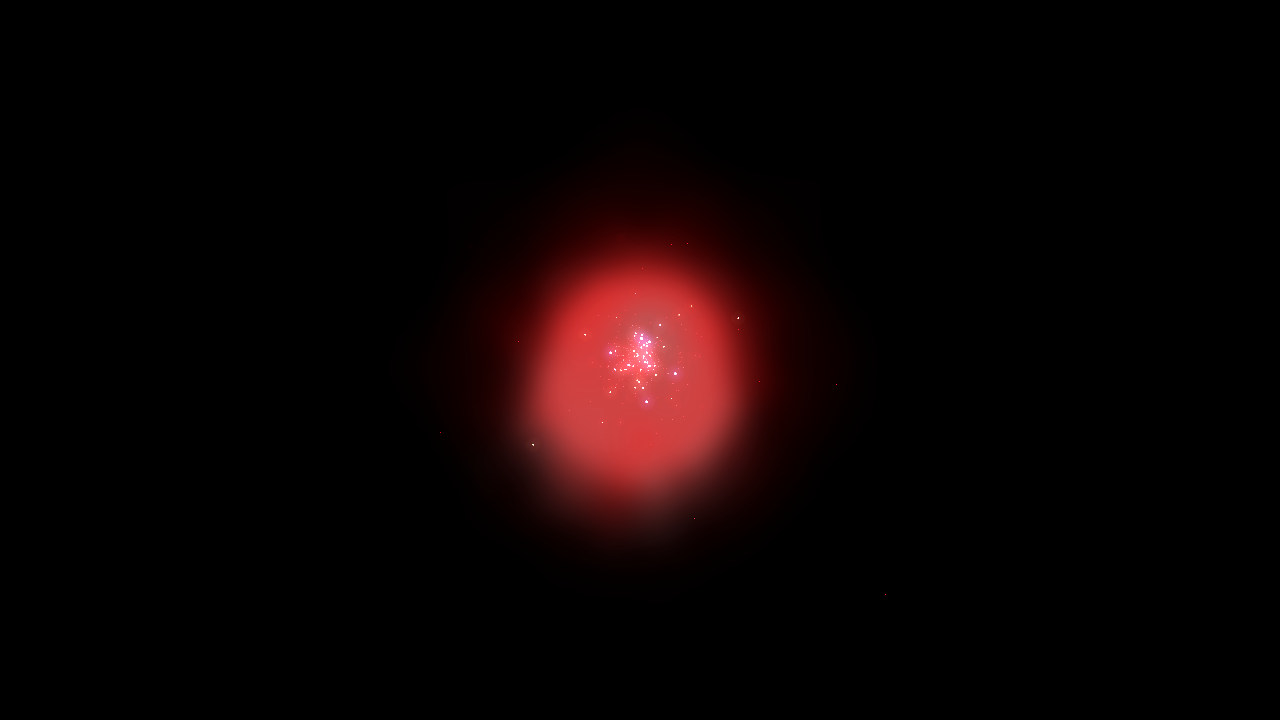}
}
\subfigure{
\includegraphics[width=\columnwidth]{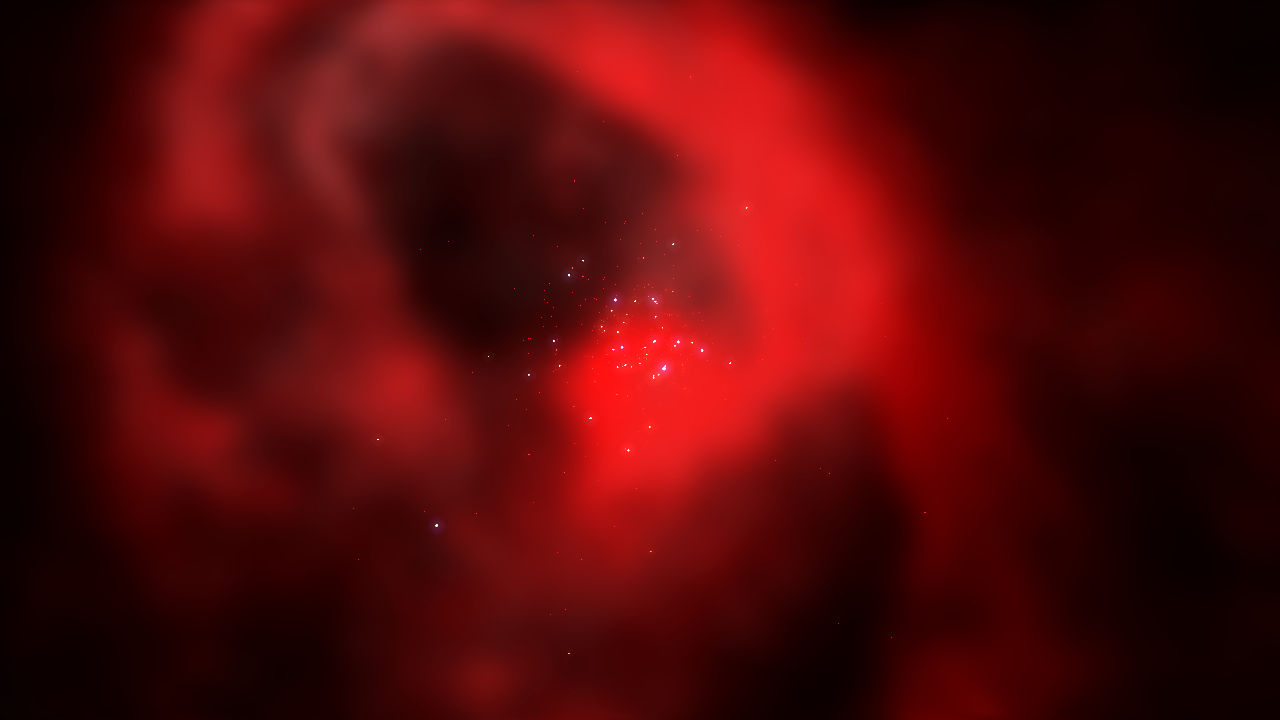}
}
\subfigure{
\includegraphics[width=\columnwidth]{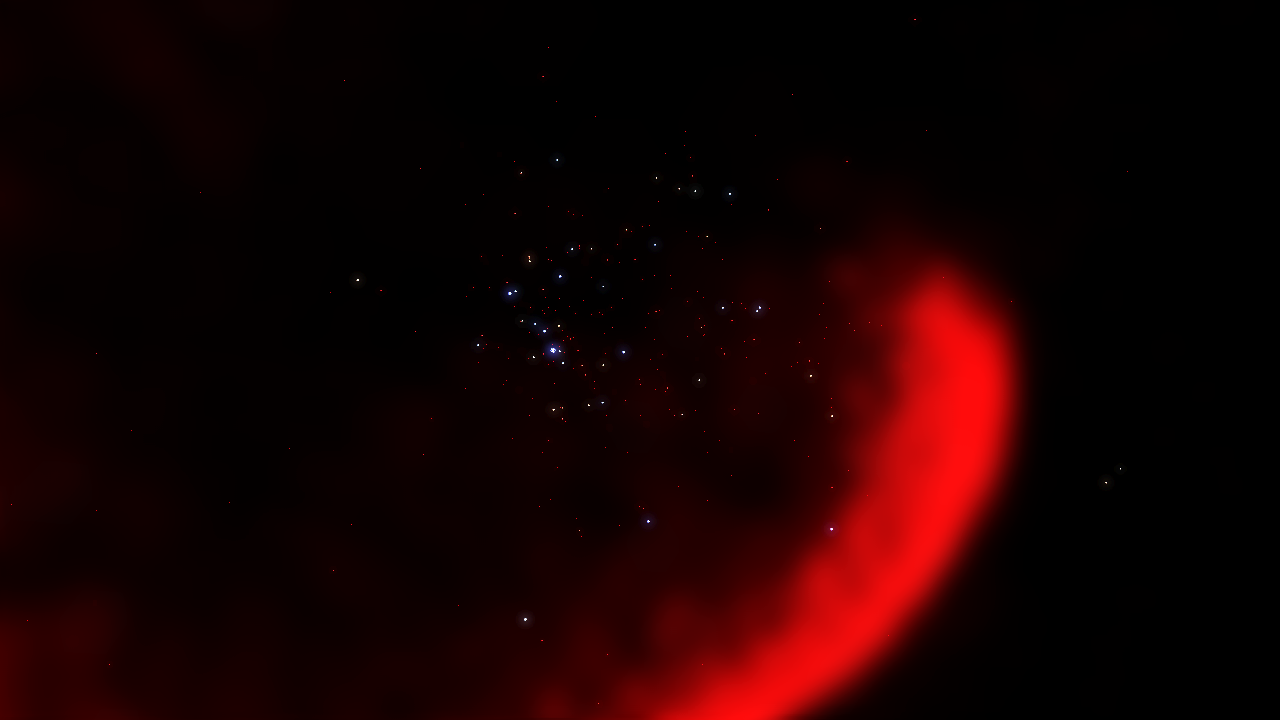}
}
\subfigure{
\includegraphics[width=\columnwidth]{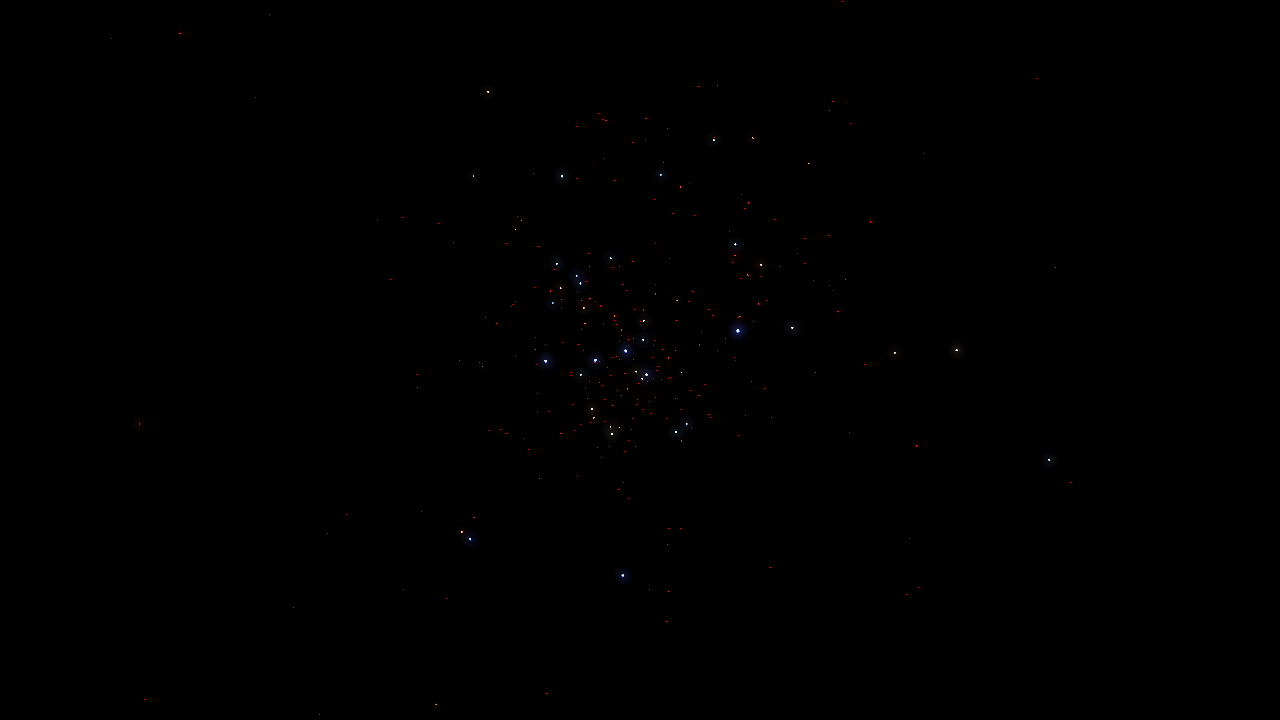}
}
\caption{3D Visualization of the embedded star cluster evolution simulation at four different times. From top to bottom: a) The initial condition, young stars embedded in a sphere of gas. b) gas is expanding. c) only a thin shell of gas around the cluster remains. d) gas completely removed from cluster (note the larger size of the cluster)}
\label{figure_simulation_output}
\end{figure}

For all our experiments, we use the same simulation: one simulating the evolution of embedded star clusters.
For details, see~\cite{stellar_evolution}.
In this simulation, an early star cluster is simulated, including the gas from which the stars formed. The stars interact with the gas, which is eventually pushed out of the cluster completely.
Also, the stars themselves evolve, leading to several of the bigger stars exploding in a supernova during the simulation.

As part of our research we have created an interactive 3D visualization of this simulation.
A video showing the output of this visualization is available on the Ibis website\footnote{see \url{http://www.cs.vu.nl/ibis/demos.html}}.
Figure ~\ref{figure_simulation_output} shows a number of screenshots of the simulation at four different stages, from the initial condition to a state where all the gas is removed completely.

\begin{figure}
\centering
\includegraphics[width=\columnwidth]{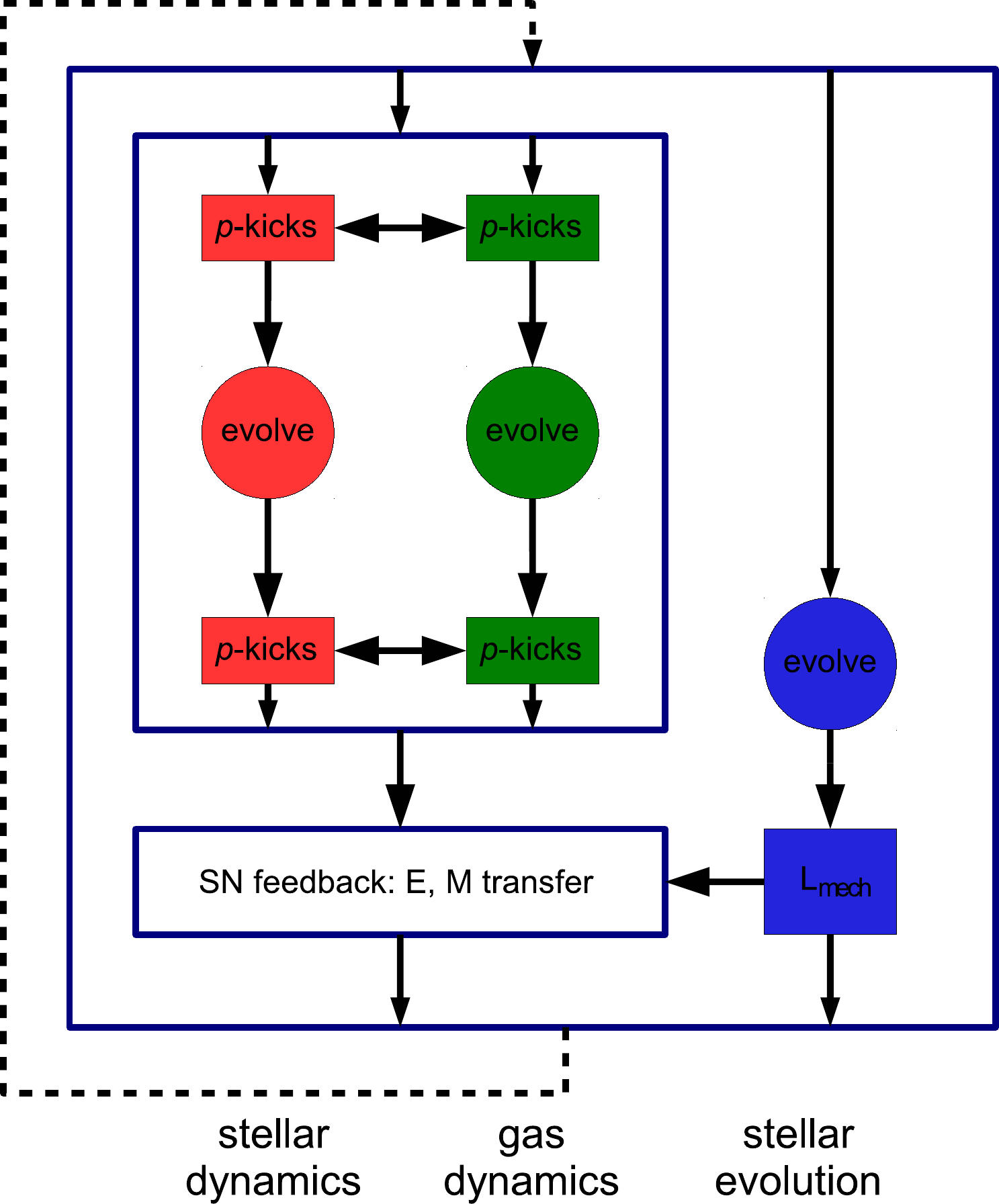}
\caption{The AMUSE~\cite{muse} gravitational/hydro/stellar evolution integrator. 
This diagram shows the calling sequence of the different AMUSE elements in
the combined gravitational/hydro/stellar solver during a time step of the 
combined solver. Circles indicate calls to the models, 
while rectangles indicate parts of the solver implemented in Python 
within AMUSE. This image taken from~\cite{stellar_evolution}.}
\label{figure_grav_gas_sse}
\end{figure}

\begin{figure}
\centering
\includegraphics[width=\columnwidth]{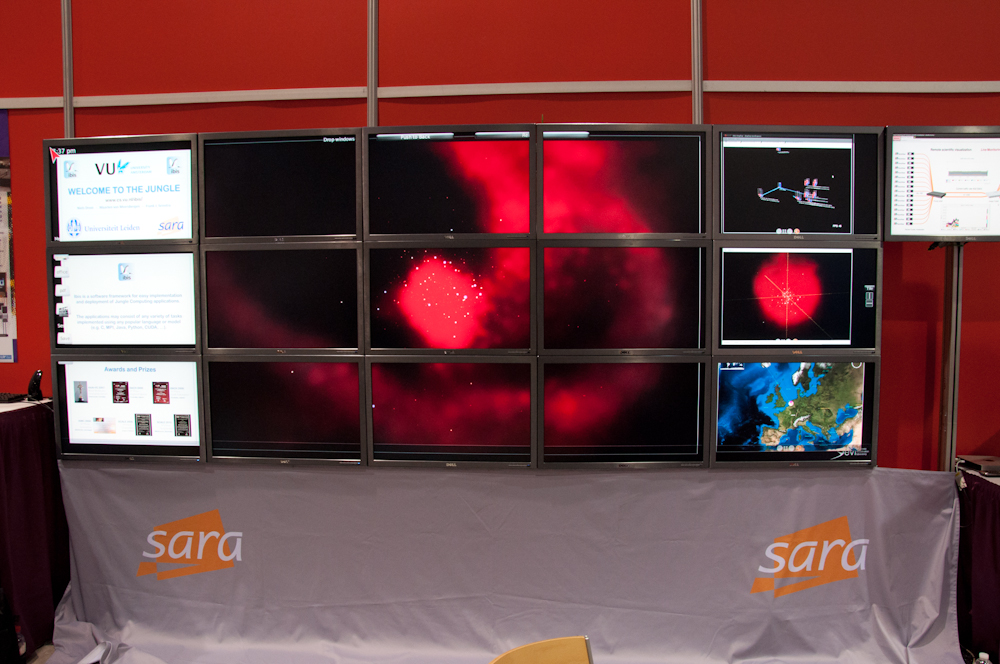}
\caption{Demonstration at SC11, Seattle USA.}
\label{figure_tpd}
\end{figure}

The embedded star cluster simulation uses four different models.
First, one simulating the gravity between the stars is used.
A number of kernels is available to do this task.
In this case we use PhiGRAPE~\cite{phigrape} (written in Fortran), which is available in both a CPU and a GPU (using CUDA) variant.
Second, the stars' evolution is simulated, using SSE~\cite{sse}.
SSE is a so-called parameterized model, which does a simple lookup of a stars' age and initial mass to determine its current state.
Since this lookup is nearly trivial, SSE is simply a sequential (Fortran) application.
Third, the gas present in the cluster is simulated using a hydrodynamics model, Gadget~\cite{gadget}.
Gadget is a CPU only model, written in C/MPI.
Last, we use a model to couple the gravity interactions between stars and gas.
The gas and gravitational models are completely independent
Since we are simulating a system with both gas and stars present, the influence between these two must be modeled explicitly as well.
For this coupling, the Octgrav~\cite{octgrav} gravitational tree model is used, implemented in C++ and CUDA.
If no GPU is available, the Fi~\cite{fi} model, written in Fortran, can be used instead.

Figure~\ref{figure_grav_gas_sse} shows a schematic view of the data flow and computation of the simulation for a single step of the simulation.
The inner box on the top left shows the gas dynamics models coupled with the gravitational (stellar) dynamics model.
The evolve step can be done in parallel, while the 'p-kicks' phases are implemented using the coupling model.
The stellar evolution model is not required at every time step of the inner coupling. It is performed at a slower rate, only exchanging state every n-th time step, depending on the parameters of the simulation~\cite{stellar_evolution}.

\subsection{SC11 Demonstration}

\begin{figure*}
\centering
\includegraphics[width=0.75\textwidth]{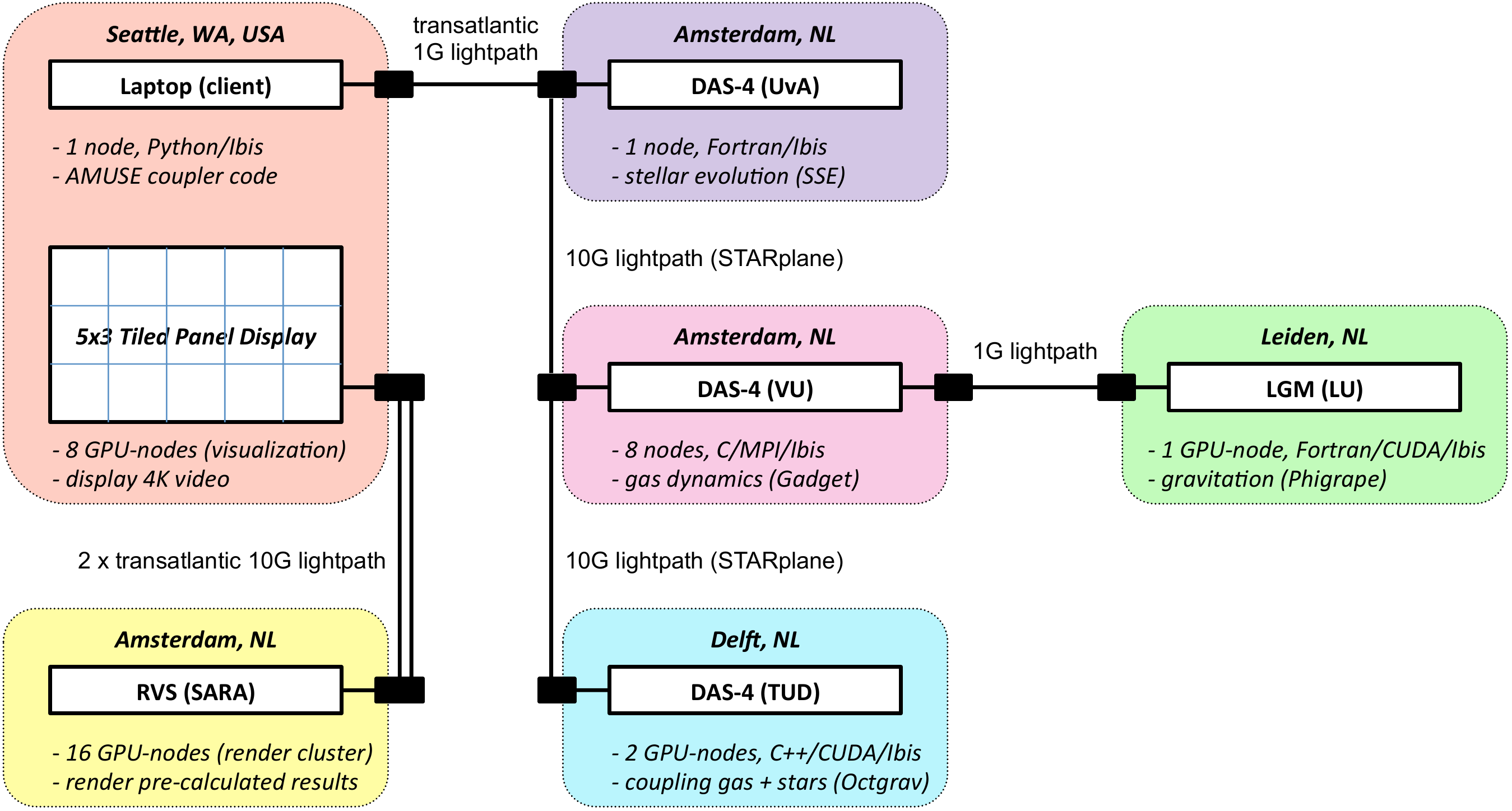}
\caption{Machines and Network used for the demonstration at SC11, Seattle, USA.}
\label{figure_SC11_setup}
\end{figure*}

We showed our Distributed AMUSE as a demonstration at the SC11 SuperComputing conference in Seattle, USA, in November 2011.
This demonstration gave us an opportunity to test our software in a Jungle environment.
We tested a worst-case scenario where the coupler was running on one side of the Atlantic ocean, and all the models were running on the other side.
Figure~\ref{figure_SC11_setup} shows the machines and network used during the demonstration.
AMUSE, and the Ibis Daemon were running on a laptop (shown on the top left).
We also used a tiled panel display to display a 4K resolution version of the 3D visualization, rendered by a 16 node cluster located in Amsterdam, The Netherlands.
Figure~\ref{figure_tpd} shows a photo of the demonstration setup.

Using a transatlantic 1G lightpath, the laptop in Seattle was connected to The Netherlands.
The stellar evolution, gas dynamics, and coupling models where run on different clusters of the DAS-4~\cite{das4}.
The DAS-4 is a distributed system consisting of 6 clusters throughout the Netherlands.
We ran each model on a different cluster.
To run the gravitational dynamics, we used the \emph{Little Green Machine}(LGM)~\cite{lgm} located at Leiden Observatory.

\begin{figure*}
\centering
\includegraphics[width=0.70\textwidth]{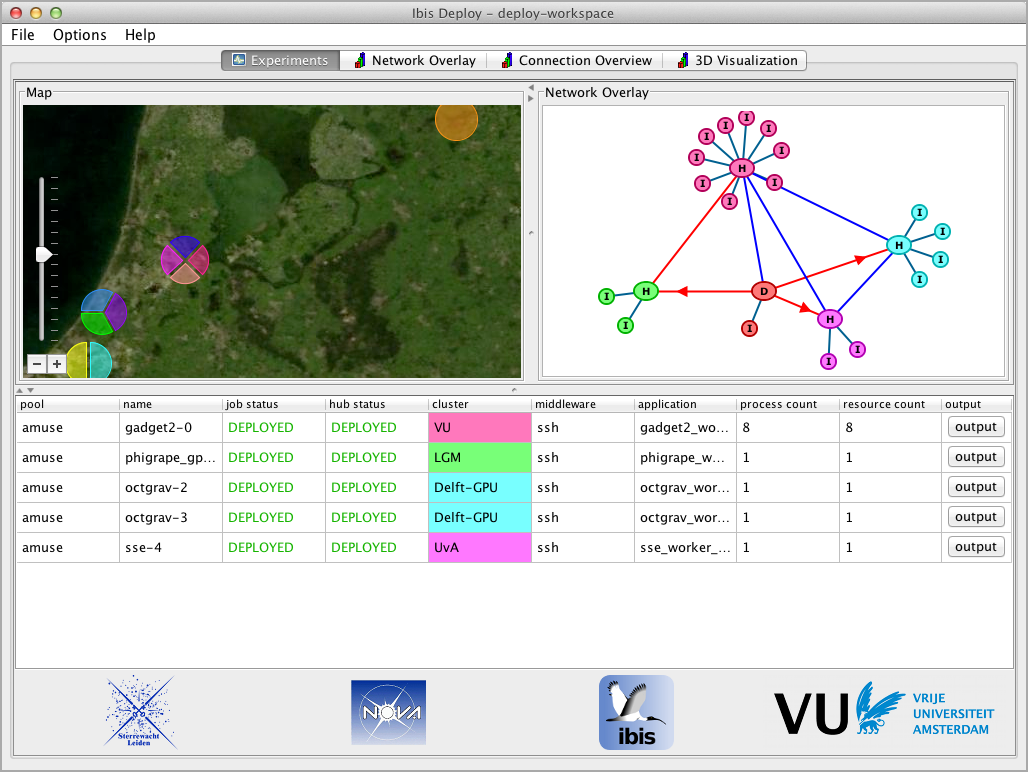}
\caption{Screenshot of IbisDeploy during Demonstration.
Top-left: available resources.
Bottom half: jobs used to start the models.
Top-right: overlay network created and used by SmartSockets to ensure connectivity.
}
\label{figure_deploy_screenshot_jobs}
\end{figure*}

As said, our prototype system comes with an optional GUI, based on the IbisDeploy GUI.
We have included a video taken of this GUI during the demonstration on our website~\footnote{see~\url{http://www.cs.vu.nl/ibis/demos.html}}.
As the video shows, the models are all started in order. See Figure~\ref{figure_deploy_screenshot_jobs} for a screenshot.
The top-left of the window shows the available resources on a map, all located in The Netherlands.
The bottom half of the window shows the different jobs used to start the models.
The top-right of the screenshot shows the overlay network created and used by SmartSockets to ensure connectivity.
Red lines denote ssh tunnels automatically setup, while arrows denote that a connection was only possible in one direction, possibly due to a firewall or NAT.

\begin{figure*}
\centering
\includegraphics[width=0.70\textwidth]{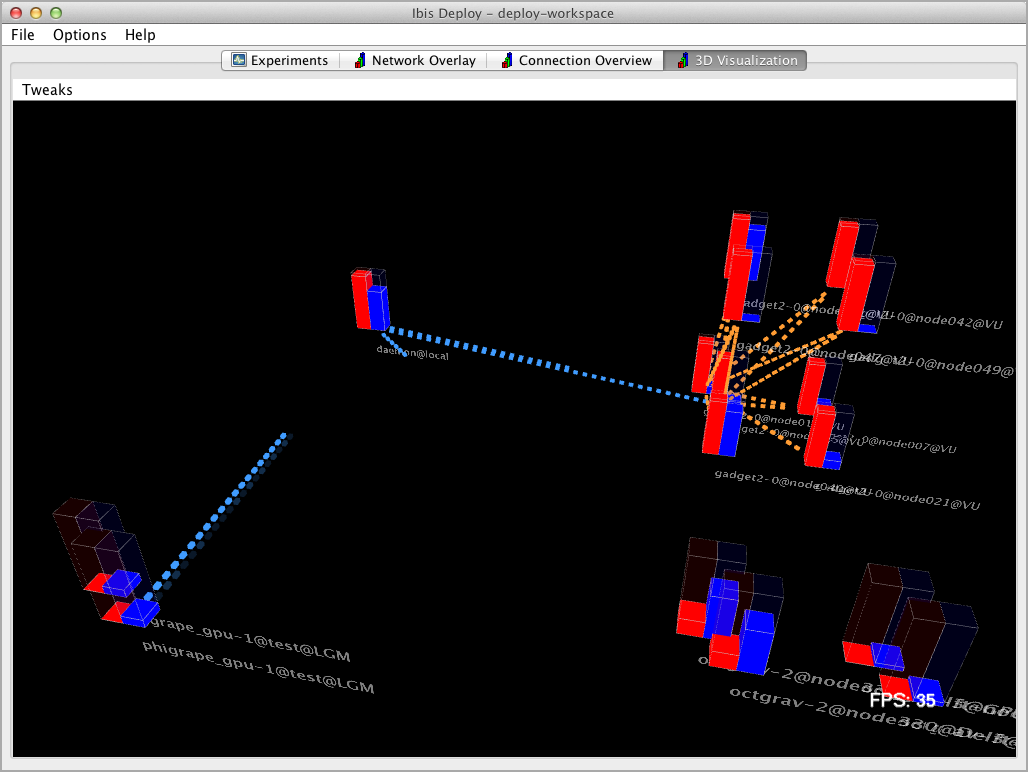}
\caption{Screenshot of IbisDeploy showing a 3D network traffic visualization.
Left-back: AMUSE on laptop (Seattle).
Left-front: phiGRAPE on LGM (Leiden).
Right-back: Gadget on 8 nodes of DAS-4 (Amsterdam).
Right-front: Octgrav on 2 nodes of DAS-4 (Delft).
Not shown: SSE on 1 node of DAS-4 (Amsterdam).
}
\label{figure_deploy_screenshot_network_viz}
\end{figure*}

Once all models are started, AMUSE starts running the simulation.
The video next switches to a number of alternative views of the system: the overlay network, all connections created in IPL, and a 3D visualization showing network traffic. See Figure~\ref{figure_deploy_screenshot_network_viz} for a screenshot.
In the left-back corner is the laptop located in Seattle.
The other locations represent the different models.
IPL traffic is shown in blue, while MPI traffic is shown in orange.
The bars at each location denote machine load (red) and memory usage (blue).
Note that the nodes running models that support GPUs have a very low load.
As the GPU is used, the CPUs in the machine are almost completely idle.

\subsection{Lab Conditions}

Unfortunately, the simulation used in our tests is too small to properly test the scalability of our software.
Still, we were able to perform some tests using a small number of resources.
First, we ran a single iteration (time step) of the simulation (the total simulation takes about 1200 iterations).
To represents a user with access to only a basic machine, we solely used the CPU in a quad core Intel Core2 machine, by selecting the Fi and phiGRAPE(CPU) models.
This setting lead to a runtime of 353 seconds per iteration.

However, suppose the user also has a GPU in his system?
Next, we also used the GeForce 9600GT in the desktop machine, by switching to Octgrav and phiGRAPE(GPU) models.
This increased performance dramatically, to 89 seconds per iteration.
This shows AMUSE is able to quickly make use of GPU resources available.

As a third scenario, suppose a user does not have a GPU for himself, but does have access to a GPU machine remotely?
We determined that the Fi coupler model was dominating the runtime in the first scenario, and ran the Octgrav model on a node of the LGM cluster in Leiden instead.
Note that we only had to change a single line in our simulation script to make this change.
Using this remote GPU instead of the local GPU increased performance even further to 84 seconds per iteration.
This is interesting, as using the compute power of a GPU 30 kilometers away is faster than using a GPU located inside our own machine.
As the LGM node we used has a Tesla C2050 GPU, this is entirely possible.
Still, the overhead of our prototype is clearly low enough for this to work.

\begin{figure*}
\centering
\includegraphics[width=0.75\textwidth]{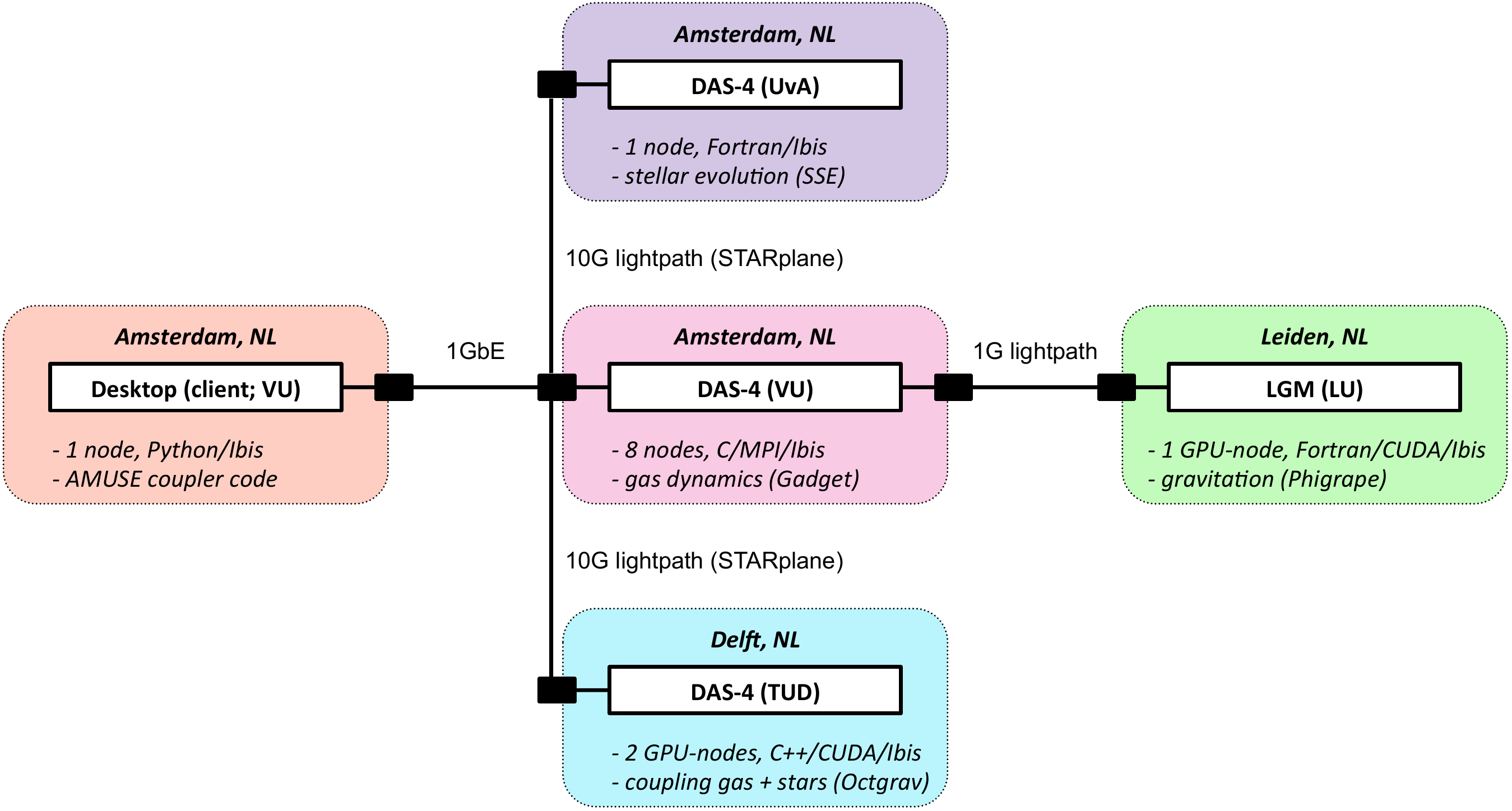}
\caption{Machines and network used for the Lab Experiments}
\label{figure_lab_setup}
\end{figure*}

Last, we tested the worst-case scenario setup of the demonstration, using the same quad core desktop instead of the laptop.
In practice, spreading all models of a small simulation across 4 different universities is normally not necessary.
This true Jungle setup increased performance even further to 62.4 seconds per iterator.
See Figure~\ref{figure_lab_setup} for the machines used in this experiment.
From these limited experiments we conclude that our software is indeed usable in a Jungle, and, even for very small problems, is capable of using remote resources effectively.

\section{Conclusions and Future Work}
\label{conclusions}

In this paper we showed a case study in Jungle Computing: running high-performance distributed \mmmks{}. We described two different \mmmks{} applications: Climate Modeling, and Computational Astrophysics. For the latter we implemented a prototype system to run simulations on a  Jungle Computing System, and showed preliminary experiments, sometimes spanning multiple continents.

The setup used to evaluate our system is small. We are currently in the process of scaling up our experiment.
Using the infrastructure that we recently acquired access to~\cite{eyr3}, including lightpaths and compute time on a number of compute resources, including a supercomputer, we plan to scale up our experiment significantly, with at least a factor 100, in the near future.

We argue that using Jungle Computing Systems is necessary to make scientific progress in these fields.
The Ibis software framework already solves many of the problems involved.
Besides showing the viability of our approach, the prototype we build is already useful in itself, and was received with enthusiasm in the computational astronomy community.

Of course, there is still a lot of room for improvement.
Fault-tolerance is an issue, and users would like to have automatic selection of appropriate resources for their simulations.
We are also planning on making a distributed version of the coupler present in AMUSE, and allow direct communication between models.
Our research in CESM is also progressing, and we hope to create a Jungle-Aware version of CESM in the near future.
We plan to continue to work on fulfilling the requirements that we listed for \mmmks{}: easy deployment, robust communication, effective monitoring, good fault-tolerance, and automatic resource discovery.

\section*{Acknowledgment}
This publication was supported by the Dutch national program COMMIT.

This work was supported by the Netherlands Research Council NWO (LGM \#643.200.503, VICI \#639.073.803 and AMUSE \#614.061.608) and by the Netherlands Research School for Astronomy (NOVA).

The authors wish to kindly thank Vianney Govers and Kees Verstoep for providing support for the LGM and DAS-4, and setting up the LGM/DAS network connection.

\bibliographystyle{abbrv}
\bibliography{paper.bib}

\end{document}